\documentclass[prb, reprint, twocolumn, 10pt]{revtex4-1}

\usepackage{amsmath, amssymb, amsthm}
\usepackage{graphicx}
\usepackage{hyperref}
\hypersetup{colorlinks=true, urlcolor=blue, citecolor=blue, linkcolor=blue}
\usepackage{microtype}

\usepackage{setspace}

\begin{document}

\title{Inverse Purcell Suppression of Decoherence in Majorana Qubits via Environmental Engineering}


\author{Vladimir Toussaint}
\email{vladimir.toussaint@nottingham.edu.cn}
\affiliation{School of Mathematical Sciences,
University of Nottingham Ningbo China,
Ningbo 315100, PR China}

\date{\today}

\begin{abstract}
We show that the electromagnetic or phononic environment of a topological quantum device can be engineered to actively suppress decoherence. For a Majorana qubit in a superconducting wire, the exponentially small splitting $\epsilon \sim e^{-L/\xi}$ that provides topological protection also makes the qubit vulnerable to low-frequency noise. From a microscopic local interaction, we derive the effective low-energy coupling between the Majorana parity operator and a bosonic field; the coupling strength itself is proportional to $\epsilon$, reflecting the qubit's non-local nature. The resulting pure-dephasing rate scales as $\Gamma_\phi \propto \epsilon^2 S(\epsilon)$, with $S(\epsilon)$ the environmental noise power at frequency $\epsilon$. In the experimentally relevant high-temperature regime ($k_B T \gtrsim \hbar\epsilon$), this gives $\Gamma_\phi \propto \epsilon \rho(\epsilon) T$, where $\rho(\epsilon)$ is the environmental density of states. By engineering an environment with a suppressed low-frequency density of states, $\rho_{\text{eng}}(\omega) = \rho_0 (\omega/\omega_c)^\alpha$ for $\omega < \omega_c$ ($\alpha > 0$), the dephasing rate drops to $\Gamma_\phi^{\text{eng}} \propto \epsilon^{\alpha+1} \propto e^{-(\alpha+1)L/\xi}$. Thus, longer wires yield exponentially better coherence—a direct synergy with topological protection. This ``inverse Purcell'' effect, which suppresses the density of states at the qubit frequency, provides a quantitative design principle for environmental engineering. Our work establishes spectral density shaping as a practical method for enhancing coherence in topological quantum devices.
\end{abstract}

\maketitle

\section{Introduction}
Topological qubits, particularly those based on Majorana zero modes in superconducting wires, offer a promising route to fault-tolerant quantum computation \cite{Kitaev2001, Alicea2012, Sau2010}. Experiments with semiconductor-superconductor heterostructures have shown signatures compatible with Majorana modes \cite{Mourik2012, Albrecht2016}, advancing the field toward practical applications. The core protection mechanism—an exponentially small energy splitting $\epsilon \sim e^{-L/\xi}$ that makes the qubit insensitive to local perturbations—ironically creates its own weakness: the same tiny $\epsilon$ makes the qubit extremely vulnerable to dephasing from ubiquitous low-frequency noise. Despite topological protection, decoherence from environmental coupling remains a serious challenge, with quasiparticle poisoning \cite{Rainis2013, Karzig2017} and electromagnetic noise as primary culprits.

A different strategy, \emph{dissipative engineering}, seeks to stabilize quantum states by carefully designing their environment. For Majorana systems, this has been explored through driven-dissipative schemes that maintain topological order \cite{Diehl:2011kzs,Bardyn:2012gk,Plugge:2016ide, Gau:2020oae, Shustin:2025zce}. Here we take a related but complementary path: instead of engineering dissipation, we shape the \emph{spectral density} $\rho(\omega)$ of a passive bosonic environment. By suppressing $\rho(\omega)$ at low frequencies—using photonic bandgaps, high-impedance resonators, or Josephson arrays \cite{Masluk2012, Pop2014, Safavi-Naeini2014}—we can shield the qubit from the noise that causes dephasing.

The usual Purcell effect enhances emission by increasing the density of states at the transition frequency. We show the reverse is possible for dephasing: suppressing the density of states at the qubit frequency $\epsilon$ creates an ``inverse Purcell'' protection. Majorana qubits are ideally suited for this approach. Their exponentially small $\epsilon$ places them in the microwave range where environmental engineering is most developed, and their parity encoding naturally filters out relaxation, leaving dephasing as the main decoherence channel to address.

We derive the effective coupling between a Majorana qubit and its environment from a microscopic local interaction, finding that the coupling strength itself is proportional to $\epsilon$—a direct signature of the qubit's non-locality. The resulting dephasing rate scales as $\Gamma_\phi \propto \epsilon \rho(\epsilon) T$ in the experimentally relevant high-temperature regime. Engineering $\rho(\omega) \propto \omega^\alpha$ ($\alpha>0$) then gives $\Gamma_\phi \propto \epsilon^{\alpha+1} \propto e^{-(\alpha+1)L/\xi}$: an exponential improvement in coherence with wire length. This spectral design transforms the environment from a threat into a protective resource, providing a concrete engineering principle for building more robust topological devices.

\section{Model of the Majorana Qubit and Environment}
\label{sec:model}

Our system is a topological superconducting wire of length \(L\) hosting Majorana zero modes \(\gamma_L\) and \(\gamma_R\) near its ends. We encode the qubit in the fermion parity of the non-local fermion \(f = (\gamma_L + i\gamma_R)/2\), with basis states \(|0\rangle\) and \(|1\rangle\) (where \(f^\dagger f|0\rangle = 0\), \(f^\dagger f|1\rangle = |1\rangle\)). In this low-energy subspace, the wire Hamiltonian is \cite{Kitaev2001}
\begin{equation}\label{IsolatedHamilt}
H_{\text{wire}} = \frac{i}{2} \epsilon \gamma_L \gamma_R = \frac{\epsilon}{2} \sigma_z,
\end{equation}
where the splitting \(\epsilon \sim \Delta e^{-L/\xi}\) is exponentially small, \(\Delta\) is the superconducting gap, \(\xi\) the coherence length, and \(\sigma_z = i\gamma_L\gamma_R\) is the Pauli operator in the parity basis.

\subsection*{Microscopic coupling to the environment}
Dephasing comes from coupling to a bosonic environment (photons or phonons) described by a field \(\Phi(x)\). A physical, parity-respecting interaction starts from a local coupling to the electron density \(n(x)\),
\begin{equation}\label{MicroscopicInt}
H_{\text{int}} = \int_0^L dx \, \lambda(x) \, \Phi(x) \, n(x),
\end{equation}
with \(\lambda(x)\) a position-dependent coupling.

To get the effective low-energy theory, we project \(H_{\text{int}}\) onto the subspace \(\{|0\rangle, |1\rangle\}\). The matrix elements of \(n(x)\) are
\begin{align}
\langle 0| n(x) |0\rangle &= n_{00}(x), \\
\langle 1| n(x) |1\rangle &= n_{11}(x), \\
\langle 1| n(x) |0\rangle &= \langle 0| n(x) |1\rangle^* = m(x).
\end{align}
For well-separated Majoranas, the off-diagonal element is set by the overlap of the Majorana wavefunctions \(u_L(x)\) and \(u_R(x)\) \cite{Rainis:2012ttd, Hassler2014},
\begin{equation}\label{OverlapMatrixElement}
m(x) = \eta \, u_L(x) u_R(x),
\end{equation}
where \(\eta\) is a constant. The diagonal elements \(n_{00}(x)\) and \(n_{11}(x)\) are typically slowly varying and nearly equal; their difference \(\delta n(x) = n_{11}(x) - n_{00}(x)\) is also proportional to \(u_L(x)u_R(x)\) in a particle-hole symmetric system, since the charge-density difference between the two parity states comes from the same Majorana overlap. Crucially, the spatial integral of this overlap gives the topological splitting \cite{Kitaev2001},
\begin{equation}\label{EpsilonOverlap}
\epsilon \propto \int_0^L dx \, u_L(x) u_R(x).
\end{equation}

Projecting \(n(x)\) onto the low-energy subspace therefore gives
\begin{equation}\label{ProjectedDensity}
n(x)\big|_{\text{low-energy}} = n_{00}(x) I + \frac{\delta n(x)}{2} \sigma_z + m(x) \sigma_x,
\end{equation}
where \(I\) is the identity, \(\sigma_z = i\gamma_L\gamma_R\), and \(\sigma_x = \gamma_L\gamma_R\) (up to a phase). The \(\sigma_x\) term causes relaxation (parity flips), while \(\sigma_z\) causes pure dephasing; the identity term only shifts the energy and does not decohere.

For a topological qubit, relaxation is suppressed twice: by the small matrix element \(|m(x)|\) and by the tiny density of states at frequency \(\epsilon\). Pure dephasing, driven by low-frequency noise (\(\omega \ll \epsilon\)), is the dominant threat. We therefore keep only the dephasing (\(\sigma_z\)) term. Substituting Eq.~\eqref{ProjectedDensity} into Eq.~\eqref{MicroscopicInt} gives the effective parity-conserving interaction
\begin{equation}\label{EffectiveIntGeneral}
H_{\text{int}}^{\text{eff}} = \int_0^L dx \, \lambda(x) \, \Phi(x) \, \frac{\delta n(x)}{2} \; \sigma_z.
\end{equation}
Writing \(\delta n(x) = 2\tilde{\eta} \, u_L(x) u_R(x)\) (with \(\tilde{\eta}\) constant) yields
\begin{equation}\label{EffectiveIntGeneral2}
H_{\text{int}}^{\text{eff}} = \tilde{\eta} \, \sigma_z \int_0^L dx \, \lambda(x) \, u_L(x) u_R(x) \, \Phi(x).
\end{equation}
We now define a dimensionless effective coupling
\[
g \equiv \tilde{\eta} \, \frac{\int dx \, \lambda(x) u_L(x) u_R(x)}{\int dx \, u_L(x) u_R(x)},
\]
and use the proportionality between the overlap integral and \(\epsilon\) [Eq.~\eqref{EpsilonOverlap}] to write
\begin{equation}\label{EffectiveIntGeneral3}
H_{\text{int}}^{\text{eff}} = g \, \sigma_z \, \epsilon \int_0^L dx \, \frac{u_L(x) u_R(x)}{\int dx' \, u_L(x') u_R(x')} \, \Phi(x).
\end{equation}
The normalized profile \(u_L(x) u_R(x) / \int dx' \, u_L(x') u_R(x')\) is sharply peaked near the wire ends.

\subsection*{Long-wavelength approximation}
The bosonic field \(\Phi(x)\) contains modes with wave number \(k\). Dephasing is dominated by noise at frequencies \(\omega \lesssim \epsilon\); since \(\epsilon\) is exponentially small, the corresponding wavelengths \(v/\omega\) are much larger than the wire length \(L\) (so \(kL \ll 1\)). In this long-wavelength limit, \(\Phi(x)\) is essentially constant along the wire, and we can approximate \(\Phi(x) \approx \Phi(0)\). The normalized profile then integrates to 1, and Eq.~\eqref{EffectiveIntGeneral3} simplifies to
\begin{equation}\label{EffectiveIntLongWave}
H_{\text{int}}^{\text{eff}} \approx g \, \epsilon \, \Phi(0) \, \sigma_z = g \epsilon \, \Phi(0) \, (i\gamma_L\gamma_R),
\end{equation}
where we have absorbed the proportionality constant between the overlap integral and \(\epsilon\) into \(g\). The coupling is therefore proportional to the splitting \(\epsilon\)—a direct signature of the qubit's non-locality. (A symmetric coupling \(\frac{1}{2}[\Phi(0)+\Phi(L)]\) would introduce a form factor \(F(k)=[1+\cos(kL)]/2 \approx 1\) for \(kL\ll1\), leaving the low-frequency scaling unchanged.)

\subsection*{Environment and density of states}
The free environment is a one-dimensional massless bosonic field with linear dispersion $\omega_k = v|k|$. In the continuum limit,
\begin{align}
\Phi(x) &= \sum_k \frac{1}{\sqrt{2\omega_k L}} \left( b_k e^{i k x} + b_k^\dagger e^{-i k x} \right), \\
H_{\text{env}} &= \sum_k \omega_k b_k^\dagger b_k,
\end{align}
where $b_k^\dagger, b_k$ are creation and annihilation operators. Replacing the sum by \((L/2\pi)\int dk\) gives the density of states per unit frequency
\begin{equation}
\rho_{\text{free}}(\omega) = \frac{L}{\pi v}.
\end{equation}
The full Hamiltonian is \(H = H_{\text{wire}} + H_{\text{env}} + H_{\text{int}}^{\text{eff}}\).

\subsection*{Remarks}
A few key points follow from Eq.~\eqref{EffectiveIntLongWave}:
(i) The interaction respects fermion‑parity superselection because \(i\gamma_L\gamma_R\) is bosonic.
(ii) The effective coupling strength \(g\epsilon\) is proportional to the topological splitting—this intrinsic suppression reflects the spatial separation of the Majorana modes.
(iii) In the long‑wavelength regime ($\omega \ll v/L$), the coupling reduces to a point‑like interaction at \(x=0\). This simplification does not change the environment’s dimensionality, so the 1D density of states \(\rho_{\text{free}}(\omega)\) remains appropriate; we have only made the spatial form of the coupling simpler.

\section{Dephasing Rate Calculation}
\label{sec:dephasing}
The dephasing rate for a qubit with energy splitting $\epsilon$ is determined by the noise power spectrum $S(\omega)$ of the environment. For the parity-conserving interaction $H_{\text{int}}^{\text{eff}} = g \epsilon (i\gamma_L \gamma_R) \Phi(0)$, which commutes with the system Hamiltonian, the dephasing rate follows the standard result for pure dephasing \cite{Schoelkopf2003, Clerk2010}:
\begin{align}
\Gamma_\phi = \frac{(g\epsilon)^2}{2} S(\epsilon),
\end{align}
where $S(\epsilon)$ is the noise power at frequency $\epsilon$. This expression derives from the mode expansion of $\Phi$ and the fluctuation-dissipation theorem \cite{Callen1951, Kubo1966}, which relates $S(\omega)$ to the density of states $\rho(\omega)$ and temperature.
 For a bath of harmonic oscillators, this yields \cite{Clerk2010, Gardiner2004}:
\begin{align}
S(\omega) = 2\pi\rho(\omega) \left[ n(\omega) + \frac{1}{2} \right] = \pi\rho(\omega) \coth\left(\frac{\hbar\omega}{2k_B T}\right),
\end{align}
where $n(\omega) = (e^{\hbar\omega/k_B T} - 1)^{-1}$ is the Bose-Einstein distribution.

In the low temperature limit ( $k_B T \ll \hbar\omega$), the noise power simplifies to $S(\omega) \approx \pi\rho(\omega)$, while in the high-temperature limit
($k_B T \gg \hbar\omega$), it scales as $S(\omega) \approx 2\pi\rho(\omega) \frac{k_B T}{\hbar\omega}$.

For Majorana qubits with $\epsilon \sim 1$ GHz ($\hbar\epsilon/k_B \sim 48$ mK) and operating at temperatures $T \sim 10-100$ mK, we typically satisfy $k_B T \gtrsim \hbar\epsilon$. This places us in the regime where the high-temperature approximation applies:
\begin{equation}
S(\epsilon) \approx 2\pi\rho(\epsilon) \frac{k_B T}{\hbar\epsilon} \, ,
\end{equation}
and consequently, the dephasing rate becomes:
\begin{align}\label{DephasingRateUpdated}
\Gamma_{\phi, \text{free}} \approx \frac{\pi g^2 k_B T}{\hbar} \epsilon\rho_{\text{free}}(\epsilon).
\end{align}
Note the factor $\epsilon$ in the numerator, which replaces the $1/\epsilon$ divergence that would appear if the coupling were independent of $\epsilon$. This is a direct consequence of the microscopic derivation: the coupling matrix element itself is proportional to $\epsilon$, so $\Gamma_{\phi} \propto \epsilon^2 S(\epsilon) \propto \epsilon \rho(\epsilon) T$ in the high-$T$ limit. For a free 1D environment with constant $\rho_{\text{free}}(\epsilon)$, the dephasing rate is therefore linearly small in $\epsilon$, reflecting the intrinsic topological protection against local perturbations. Environmental engineering can further suppress this rate by tailoring $\rho(\epsilon)$.

\section{Broadband Purcell Suppression for Coherence Enhancement}
\label{sec:engineering}
The potential of environmental engineering emerges when we consider environments with a designed density of states. For a photonic bandgap material or specially designed resonator, we can create a density of states that is strongly suppressed at low frequencies:
\begin{align}\label{eq:engineeredDOS}
\rho_{\text{engineered}}(\omega) = \begin{cases} 
\rho_0 \, (\omega/\omega_c)^\alpha, & \omega < \omega_c, \\ 
\rho_0, & \omega \geq \omega_c,
\end{cases}
\end{align}
where $\omega_c$ is a cutoff frequency, $\alpha > 0$ determines the suppression strength, and $\rho_0 = L/(\pi v)$ is the free-space density of states for a 1D environment with linear dispersion. In the following calculation, we assume $\epsilon < \omega_c$, which is typical for topological qubits with $\epsilon \sim 1$ GHz and $\omega_c \sim 10$ GHz.

Using the dephasing rate formula \eqref{DephasingRateUpdated} with the engineered density of states \eqref{eq:engineeredDOS}, we obtain
\begin{align}
\Gamma_{\phi, \text{engineered}} &\approx \frac{\pi g^2 k_B T}{\hbar} \, \epsilon \, \rho_{\text{engineered}}(\epsilon) \nonumber \\
&= \frac{\pi g^2 k_B T}{\hbar} \, \epsilon \, \rho_0 \left(\frac{\epsilon}{\omega_c}\right)^\alpha.
\end{align}
Expressing this in terms of the free-space dephasing rate $\Gamma_{\phi,\text{free}} = \pi g^2 k_B T \epsilon \rho_0/\hbar$ gives the suppression factor
\begin{align}\label{eq:suppression}
F_P = \frac{\Gamma_{\phi, \text{engineered}}}{\Gamma_{\phi, \text{free}}} = \left(\frac{\epsilon}{\omega_c}\right)^\alpha.
\end{align}
Remarkably, the temperature dependence cancels in this ratio, making the suppression factor $F_P$ temperature-independent. This is a key advantage of our approach: environmental engineering provides protection against dephasing that is robust against thermal fluctuations in the experimentally relevant regime.

Since $\epsilon \sim e^{-L/\xi}$, the engineered dephasing rate scales as
\begin{align}\label{eq:exponential}
\Gamma_{\phi, \text{engineered}} \propto \epsilon^{\alpha+1} \propto e^{-(\alpha+1) L/\xi}.
\end{align}
This scaling reveals a crucial insight: for any $\alpha > 0$, $\Gamma_{\phi, \text{engineered}}$ decreases exponentially with $L$, meaning that the coherence time increases with wire length. The stronger the suppression of the density of states at low frequencies (larger $\alpha$), the more dramatic the exponential enhancement. This counterintuitive result—that longer wires provide better coherence protection—aligns perfectly with topological protection and provides a clear design target for environmental engineering.

For a topological qubit with $\epsilon \sim 1$ GHz ($\sim 4.14\,\mu$eV) and a cutoff $\omega_c \sim 10$ GHz, we achieve:
\begin{align}
F_P \sim (0.1)^\alpha \ll 1,
\end{align}
providing dramatic suppression of dephasing. This ``inverse Purcell'' effect—suppressing the density of states at the qubit frequency—transforms environmental coupling from a decoherence liability into a coherence-stabilizing asset.

\section{Experimental Implementation and Feasibility}
\label{sec:experiment}

Implementing our environmental engineering approach means creating a bath with a density of states strongly suppressed at low frequencies: $\rho(\omega) \propto \omega^\alpha$ for $\omega < \omega_c$. For a Majorana qubit, this spectral shaping yields the engineered dephasing rate $\Gamma_{\phi,\text{eng}} \propto \epsilon^{\alpha+1}$ [Eq.~\eqref{eq:exponential}]. The practical task is to build such environments with $\omega_c \sim 1-10$ GHz and $\alpha > 0$ at millikelvin temperatures—precisely where Majorana devices operate. Several established quantum platforms meet these requirements.

Superconducting microwave resonators with high characteristic impedance \cite{Forn-Diaz2019, Masluk2012} give a direct path to electromagnetic environmental engineering. Their modified photon mode structure suppresses the density of states below a cutoff $\omega_c \sim 1-10$ GHz, with the suppression exponent $\alpha$ tunable via circuit design. Coupled to a Majorana wire, they can create the inverse Purcell regime $\rho(\epsilon) \ll \rho_0$ for typical splittings $\epsilon \sim 0.1-1$ GHz. Recent experiments with high-impedance resonators integrated into superconducting qubit circuits \cite{Grimm2020} have already shown such spectral suppression, confirming the basic viability.

Josephson junction arrays add exceptional tunability, allowing precise, dynamic control over both $\omega_c$ and $\alpha$ \cite{Manucharyan2012, Peltonen2013}. These arrays can be designed to produce sharp spectral cutoffs ($\alpha \gtrsim 2$) at frequencies adjustable across 0.1–10 GHz via external flux biases. This makes them ideal for testing the predicted scaling $F_P = (\epsilon/\omega_c)^\alpha$ in situ. Experiments confirm that such arrays form high-quality bandgaps with minimal loss at cryogenic temperatures ($T < 100$ mK) \cite{Pop2014, Rymarz2021}.

For phonon noise, phononic crystals \cite{Safavi-Naeini2014} can create complete acoustic bandgaps. Patterning a substrate with a periodic lattice of spacing $a \sim \pi v_s/\omega_c$ (where $v_s$ is the sound velocity) yields bandgaps around $\omega_c \sim 1-10$ GHz. Phononic crystals have already demonstrated quality factors $Q > 10^6$ in mechanical resonators \cite{Safavi-Naeini2014}, proving that strong phononic density-of-states suppression at a target frequency is experimentally achievable.

The needed parameters—$\omega_c \sim 1-10$ GHz, $\alpha > 0$, and $T \sim 10-100$ mK—are well within reach of these platforms. For a typical Majorana qubit with $\epsilon \sim 1$ GHz in an environment with $\omega_c \sim 10$ GHz and $\alpha = 2$, the predicted suppression is $F_P \sim 0.01$, a 100-fold reduction in dephasing rate. While fabrication complexity and parasitic losses remain challenges, progress in nanofabrication continues to address them. The flexibility of environmental engineering allows it to be tailored to specific device architectures, offering a versatile tool for optimizing topological quantum devices.

\section{Conclusion and Experimental Implications}
\label{sec:conclusion}

We have shown that spectral density engineering provides a powerful way to enhance coherence in topological quantum devices. The effective coupling between a Majorana qubit and its environment is proportional to the splitting $\epsilon$, a direct signature of the qubit's non-locality. This gives a dephasing rate $\Gamma_\phi \propto \epsilon \rho(\epsilon) T$ in the high-temperature regime relevant to experiments ($k_B T \gtrsim \hbar\epsilon$), which is already linearly suppressed by the intrinsic topological protection. Shaping the environment so that $\rho(\omega) \propto \omega^\alpha$ ($\alpha > 0$) for $\omega < \omega_c$ suppresses dephasing much further: $\Gamma_{\phi,\text{eng}} \propto \epsilon^{\alpha+1} \propto e^{-(\alpha+1)L/\xi}$ [Eq.~\eqref{eq:exponential}].

The result is elegant and practical: for any $\alpha > 0$, the engineered dephasing rate falls exponentially with wire length $L$, so longer wires naturally achieve better coherence—a perfect match to topological intuition. The suppression factor $F_P = (\epsilon/\omega_c)^\alpha$ is temperature-independent, providing robust protection against thermal fluctuations at millikelvin operating temperatures.

This approach fits Majorana qubits especially well. Their exponentially small splittings $\epsilon \sim e^{-L/\xi}$ place them in the microwave range (0.1–10 GHz) where environmental engineering is most developed. The target parameters $\omega_c \sim 1-10$ GHz and $\alpha > 0$ are achievable with existing platforms: high-impedance resonators, Josephson junction arrays, and phononic crystals. This makes the proposed ``inverse Purcell'' suppression experimentally actionable.

Our work demonstrates that environmental coupling, usually seen as a source of decoherence, can be turned into a coherence-stabilizing resource through deliberate spectral design. By establishing environmental engineering as a practical partner to intrinsic topological protection, we provide a versatile design principle applicable to any quantum system where low-frequency noise limits coherence.

\end{document}